\documentclass{article}
\usepackage[utf8]{inputenc}
\usepackage{amsmath}
\usepackage{graphicx}
\usepackage{hyperref}
\usepackage{geometry}
\usepackage{listings}
\usepackage{float}
\usepackage[numbers]{natbib}
\title{Decoupling Identity from Access: Credential Broker Patterns for Secure CI/CD}
\author{Surya Teja Avirneni \\
Cloud Platform and Security Architect \\
IEEE Member, ISC2 Certified, ACM Member \\
United States}
\date{April 2025}

\begin{document}

\maketitle

\begin{abstract}
In modern CI/CD systems, workload identity alone is not sufficient to enforce secure access. While SPIFFE provides runtime-issued, verifiable identities for ephemeral jobs and workloads, these identities must be mapped to actual access rights\textemdash often cloud roles, secrets, or APIs. Without a layer in between, the identity-to-access flow becomes tightly coupled, brittle, and hard to govern.

This paper introduces the concept of a credential broker: a policy-aware, runtime mediator that decouples identity from access in CI/CD pipelines. We outline patterns where SPIFFE IDs are evaluated by brokers such as OPA or Cedar before issuing cloud credentials or internal tokens. This enables just-in-time, least-privilege access and supports human approvals, context-based gating, and audit visibility.

The broker model eliminates static role bindings, supports zero standing privilege, and introduces a clean separation of concerns between identity issuance and access fulfillment. We conclude with a reference architecture and preview how these brokered decisions can evolve into intent-aware governance loops in Part 3 of this series.

Recent guidance from the NSA and CISA underscores that CI/CD pipelines are high-value targets for malicious actors and recommends adopting Zero Trust principles—explicitly advocating for runtime access controls, audit logging, and the minimization of long-lived credentials~\cite{nsa-cisa-cicd}.

\end{abstract}

\section{Introduction}

Continuous Integration and Continuous Delivery (CI/CD) pipelines are now foundational to modern software delivery, but they have also become high-value targets for attackers seeking to exploit gaps between identity and access~\cite{nsa-cisa-cicd}. In our previous paper, \textit{Establishing Workload Identity for Zero Trust CI/CD: From Secrets to SPIFFE-Based Authentication}, we demonstrated how ephemeral jobs—such as GitHub Actions or Kubernetes workloads—can be issued verifiable, runtime identities using SPIFFE. This shift from static secrets to attested identities marks a major step toward Zero Trust, enabling workloads to prove who they are based on platform and environment attributes rather than hardcoded tokens.

However, identity alone is not enough. A SPIFFE ID by itself does not entitle a CI job to deploy to production, push to a container registry, or access sensitive databases. The critical missing layer is a runtime policy gate that evaluates not just identity, but also context, intent, and explicit approvals. This is especially urgent as non-human identities (NHIs)—automated jobs, bots, and service accounts—now outnumber human users in enterprise environments and are projected to grow 10x by 2026~\cite{gartner-nhi}.

In many organizations, the link between identity and access is still managed through static IAM role mappings or manual credential distribution. While this may suffice in simple, single-team setups, it quickly breaks down at scale. The result is over-permissioned roles, brittle trust boundaries, and limited auditability—problems highlighted in recent NSA and CISA guidance, which calls for runtime access controls and the minimization of long-lived credentials~\cite{nsa-cisa-cicd}.

As CI/CD pipelines become more dynamic and decentralized, it is clear that identity and access must be decoupled. Identities should be portable and verifiable; access should be evaluated in real time, based on policy, context, and justification.

This paper introduces the concept of a \textit{credential broker}: a policy-aware, runtime mediator that sits between workload identity and access. The broker evaluates a workload’s SPIFFE ID, consults policy engines such as OPA or Cedar, and issues scoped, short-lived credentials—such as cloud tokens or database keys—only when policy conditions are met. This enables just-in-time access, zero standing privilege, and fine-grained audit trails, while simplifying policy authoring and enforcement.

In the sections that follow, we present core broker patterns, reference architectures, and real-world implementation blueprints. We show how this model not only addresses today’s governance gap, but also lays the foundation for intent-aware and continuous authorization in the next phase of Zero Trust CI/CD.

\section{The Governance Gap}

\begin{figure}[h]
  \centering
  \includegraphics[width=0.4\textwidth]{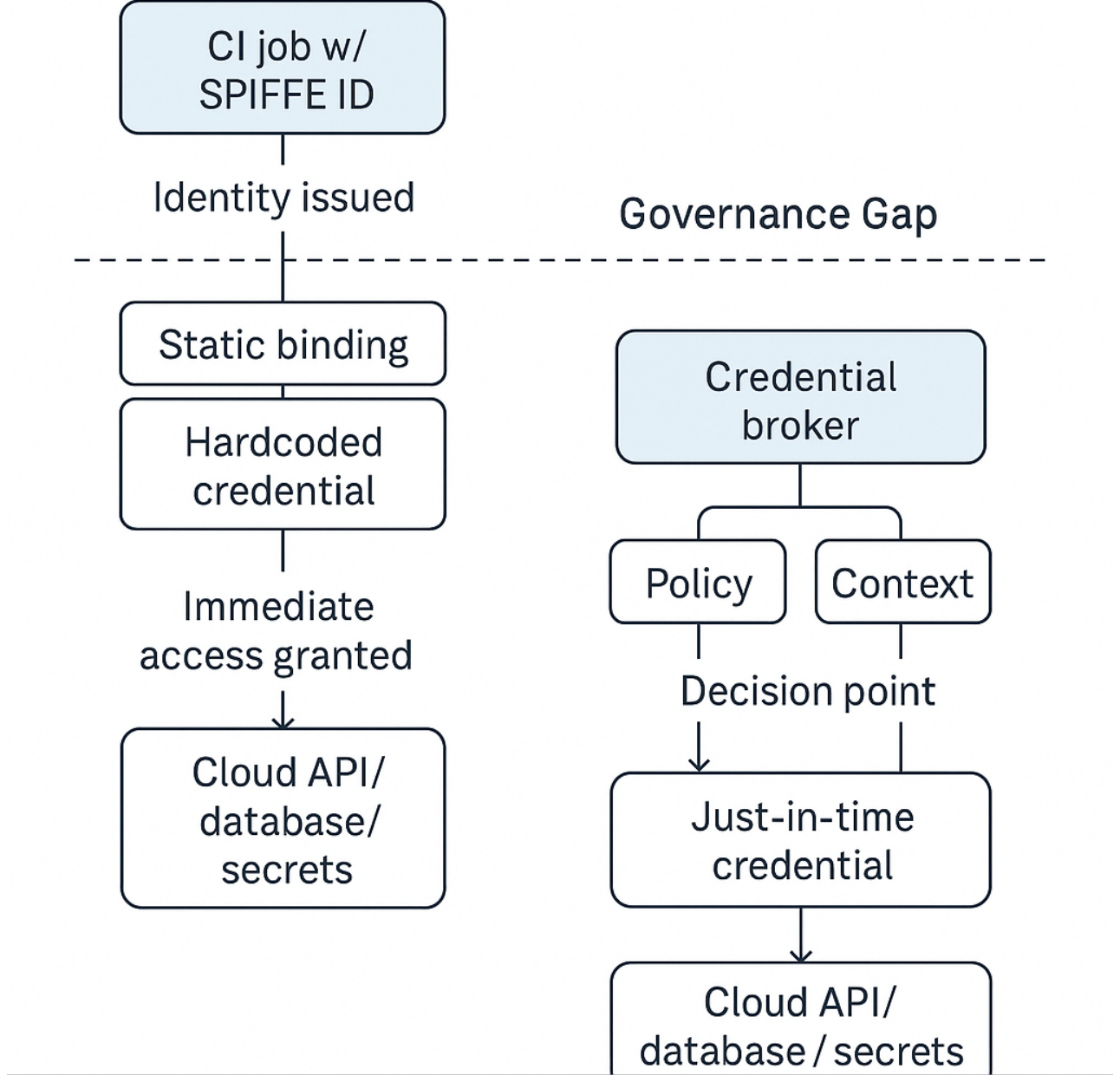}
  \caption{The governance gap between identity issuance and access enforcement. Without a broker, access is either statically bound or manually injected. A credential broker introduces a runtime decision point to enforce policy before issuing credentials.}
  \label{fig:governance-gap}
\end{figure}

Most CI/CD systems today have a gap between identity and access. Even when jobs are issued strong identities—like SPIFFE IDs or OIDC tokens—those identities are either implicitly trusted or mapped to access through static policy bindings. This introduces a class of architectural risks that show up more clearly as organizations grow.

In small setups, a pipeline might fetch cloud credentials using an identity token, and everything works. But in enterprise environments with shared platforms and dozens of teams, the assumptions start to break. Teams may not know which workloads can access which credentials. Secrets may be scoped too broadly. IAM roles might be shared across environments. Access decisions are no longer transparent.

The root of the problem is coupling. When identity issuance, credential injection, and policy enforcement all happen in the same place—or worse, are hardcoded together—it becomes difficult to understand and govern access across environments.

This challenge is consistent with NIST SP 800-204 guidance, which highlights the need for runtime authorization controls in microservices and containerized systems \cite{nist204}.

What’s missing is a runtime decision point—a control plane layer where access is granted based on identity, context, and intent. This is not about authentication alone. It’s about answering questions like:

\begin{itemize}
  \item Which workload is making this request?
  \item Why is it requesting access?
  \item Should access be granted at this time, under current policy?
\end{itemize}

Without this decision point, organizations fall back to static IAM bindings, long-lived credentials, or per-pipeline exceptions. The result is credential sprawl, policy drift, and limited audit visibility.

Credential brokers fill this gap. They act as intermediaries between identity and access, issuing credentials only when policy conditions are met. These brokers evaluate runtime context, policy, and intent before issuing short-lived credentials.

Before diving into broker patterns and deployment topologies, we take a closer look at common anti-patterns that arise when access is handled without a dedicated control point.

\section{Common Anti-Patterns in Access Handling}

When CI/CD systems lack a centralized access broker, teams often resort to brittle workarounds. These patterns may work at small scale, but they quickly break down in enterprise environments:

\begin{itemize}
  \item \textbf{Inline Credential Injection:} Jobs are preloaded with cloud access keys or secrets, often using environment variables or init scripts. These secrets are hard to rotate and may persist in logs or caches.
  
  \item \textbf{Static Role Mapping:} Identities (e.g., SPIFFE IDs or GitHub OIDC tokens) are mapped directly to IAM roles or service accounts with fixed permissions. The lack of runtime policy means access is always granted, regardless of context.

  \item \textbf{Global Secrets Mounts:} CI runners are configured to mount vault paths or secrets volumes that apply to all jobs, regardless of intent or environment. This breaks least privilege.

  \item \textbf{Approval by Convention:} Teams rely on manual Slack approvals or comments in PRs to signal that access is okay—without tying those approvals to actual authorization decisions.

  \item \textbf{Cross-Environment Reuse:} The same identity or secret is used across dev, staging, and production. Compromise in one environment leads to lateral movement.
\end{itemize}

These anti-patterns persist because there’s no clear separation between identity and access, or no runtime gate that evaluates justification. A well-placed credential broker interrupts this pattern: it provides a single control point for deciding if, when, and how credentials should be issued.

\section{Credential Broker Design Patterns}

\subsection{Why SPIFFE Alone Isn’t Enough}

A common question arises when discussing credential brokers: if SPIFFE already supports attestation and context-aware identity issuance, why not use it directly for access control?

The answer lies in the scope and design philosophy of SPIFFE. SPIFFE provides \textbf{identity}, not access. While it offers powerful attestation mechanisms—such as verifying container hashes, Kubernetes metadata, or cloud tags—and can include contextual claims in JWT-SVIDs, it does not make access decisions or issue downstream credentials.

SPIFFE provides verifiable workload identity, but does not itself enforce access control or issue downstream credentials~\cite{spiffe-concepts}.

\textbf{What SPIFFE does:}
\begin{itemize}
  \item Attests workload identity at runtime via the SPIRE Agent.
  \item Issues cryptographically verifiable identity documents (SVIDs).
  \item Includes contextual attributes in SVIDs (especially JWTs).
\end{itemize}

\textbf{What SPIFFE does \textit{not} do:}
\begin{itemize}
  \item Grant or deny access to cloud roles, databases, or secrets.
  \item Evaluate external policies (e.g., who can access which resources).
  \item Rotate credentials, handle expiration, or log access events.
\end{itemize}

\textbf{This is where credential brokers come in.} A broker serves as the runtime \textit{decision point} and \textit{credential issuer}. It receives a workload’s SPIFFE ID (and possibly a JWT-SVID with claims), evaluates whether access should be granted based on policy and context, and then issues a time-bound credential. This could be an AWS STS token, a short-lived DB cert, or a Vault secret. Brokers can also integrate approval workflows and audit trails—functions outside the scope of SPIFFE.

\textbf{Analogy:} SPIFFE is a passport authority; the broker is border control. Your SPIFFE ID proves who you are, but the broker determines if you're allowed into the system and under what conditions.

In this architecture, SPIFFE and credential brokers are complementary: SPIFFE establishes a secure, verifiable identity layer; the broker operationalizes policy-driven access using that identity.

Once a governance gap is identified, the next step is to close it with a control point that makes runtime decisions based on policy and context. Credential brokers serve this role. They are not identity providers themselves, but instead act as intermediaries that mint short-lived credentials based on verified identity and policy checks.

A broker sits between the CI job and the target service (such as a cloud API or database). It receives a workload identity—often in the form of a SPIFFE JWT-SVID—and evaluates it alongside request metadata, runtime context, and access policy. If all checks pass, the broker issues a credential, scoped to the specific operation and limited in time. This ensures that access is granted only when justified, and that no static credentials are embedded in job configurations or environments.

We now describe three practical design patterns that build on this model.

\subsection{Broker-in-the-Middle}

In this pattern, the broker is colocated with the workload—often as a sidecar, local agent, or ephemeral container. It receives workload identity via the SPIRE Workload API and uses that to generate or fetch a just-in-time credential. Popular implementations include Vault Agent or custom token generators that call AssumeRoleWithWebIdentity using the SPIFFE-issued JWT-SVID.

This model keeps the broker close to the job environment, minimizing latency and enabling tight coupling with job runtime context.

\subsection{Policy-Gated Access}

Policy engines such as Open Policy Agent (OPA) enable fine-grained, attribute-based access control (ABAC) by evaluating workload identity, resource attributes, and environmental context~\cite{opa-abac}.

Here, the broker integrates with a policy engine—like Open Policy Agent (OPA) or Cedar—to evaluate whether a workload’s identity and metadata are sufficient to authorize access. The broker does not just issue credentials based on identity alone; it makes decisions based on rules like:

\begin{itemize}
  \item Is this CI job allowed to access production?
  \item Was approval recorded for this deployment?
  \item Is the access justified under current SLA conditions?
\end{itemize}

This pattern separates identity and authorization, enabling strong audit and justification-aware gating of sensitive actions.

\subsection{Just-in-Time Tokenization}

Cloud-native services such as AWS STS exemplify the principle of issuing temporary, scoped credentials to minimize risk and support least-privilege access~\cite{aws-sts}.

In this model, the broker issues ephemeral credentials—such as STS tokens, short-lived API keys, or scoped session tokens—only when policy allows. These credentials are tied to the job identity, valid for minutes, and not reusable outside of the approved scope.

This ensures that even if a token is leaked, its usefulness is severely limited. It also avoids permanent IAM role assumptions or cross-tenant over-permissioning.

Together, these patterns form the building blocks for Zero Trust access in CI/CD: dynamic credentials, runtime decisions, and policy-aligned authorization.

\section{Reference Architectures and Deployment Models}

\begin{figure}[H]
  \centering
  \includegraphics[angle=270, width=0.5\textwidth]{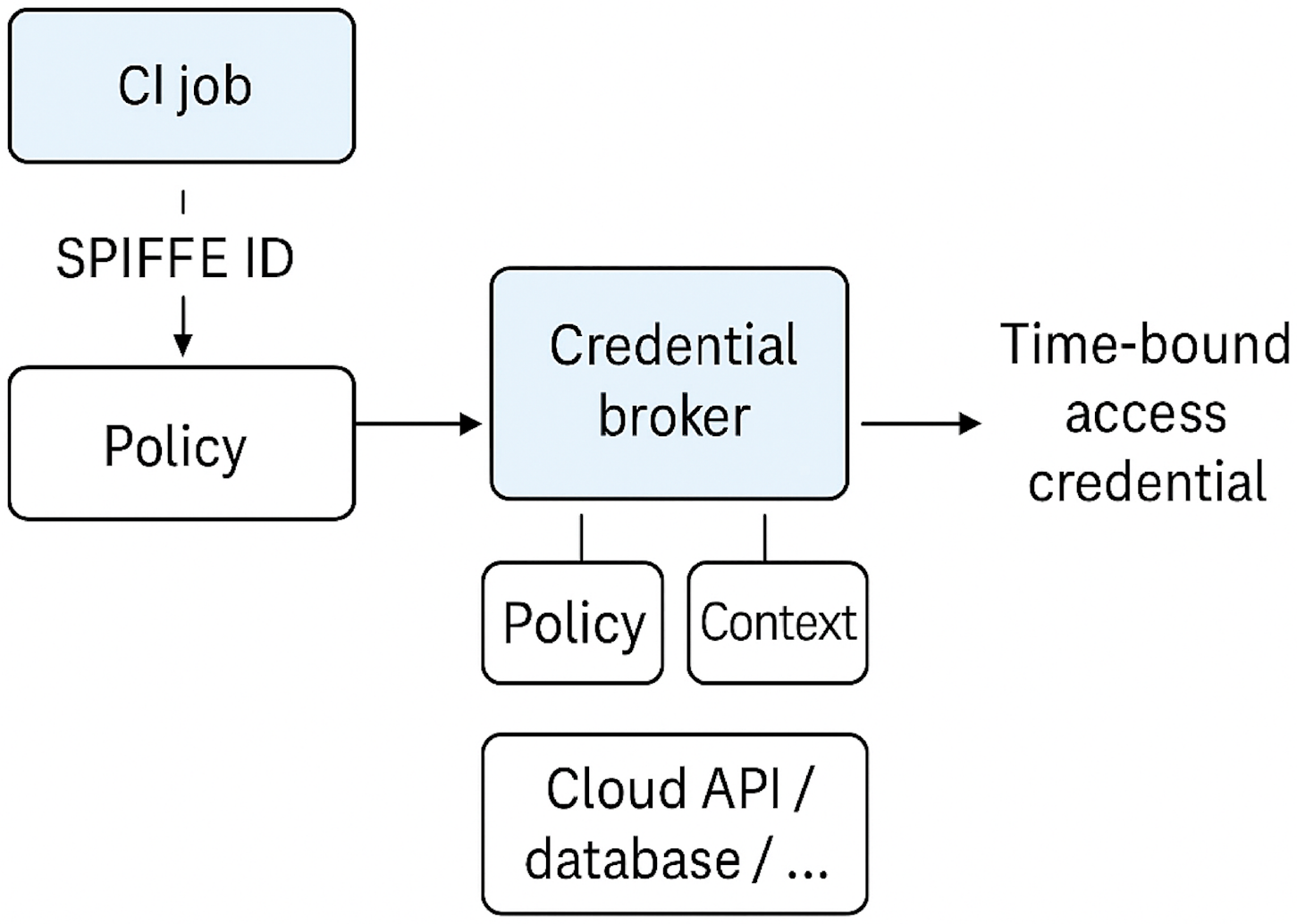}
  \caption{Reference architecture showing a SPIFFE-authenticated CI job interacting with a credential broker that evaluates policy and issues time-bound access credentials.}
  \label{fig:broker-architecture}
\end{figure}

Credential brokers can be deployed in multiple configurations depending on organizational structure, cloud environment, and CI/CD architecture. Below we describe three reference models that reflect real-world usage patterns.

\subsection{Self-Hosted Brokers}

In this model, each CI job (e.g., GitHub Actions self-hosted runner or Kubernetes Job) communicates with a local credential broker that validates its SPIFFE ID and issues credentials scoped to the job's identity. The broker is deployed alongside the job or within the same namespace.

\textbf{Advantages:} 
\begin{itemize}
  \item Strong isolation per job or namespace
  \item Lower blast radius for credential exposure
  \item Flexibility in platform-specific attestors
\end{itemize}

\textbf{Challenges:}
\begin{itemize}
  \item Operational overhead to deploy and manage brokers per job or cluster
  \item Limited cross-job policy visibility
\end{itemize}

\subsection{Centralized Broker Service}

A shared broker service acts as the control point for all CI/CD jobs across the organization. It receives SPIFFE- or OIDC-based identity assertions, consults a policy engine (e.g., OPA or Cedar), and issues short-lived credentials or tokens.

\textbf{Advantages:} 
\begin{itemize}
  \item Centralized audit and logging
  \item Uniform policy enforcement across teams
  \item Easier integration with enterprise secrets stores (e.g., Vault, STS, database proxy)
\end{itemize}

\textbf{Challenges:}
\begin{itemize}
  \item Higher availability requirements
  \item Broker becomes a sensitive dependency
\end{itemize}

\subsection{Federated and Multi-Cloud Scenarios}

In organizations operating across clouds or trust domains, SPIFFE federation enables brokers in each domain to recognize identities issued by trusted peers. This enables CI jobs in one domain (e.g., GitHub Actions) to request credentials from brokers operating in another (e.g., AWS, GCP).

\textbf{Advantages:}
\begin{itemize}
  \item Seamless integration across cloud boundaries
  \item Domain-scoped authorization using trust bundles
\end{itemize}

\section{Design Considerations and Limitations}

While credential brokers offer strong architectural advantages, they also introduce design tradeoffs. Implementing these systems at scale requires careful planning around latency, trust boundaries, and integration with existing cloud identity systems.

\subsection{Trust Anchors and Broker Placement}

Credential brokers must be deployed in a way that ensures trustworthiness. If the broker is compromised or misconfigured, it becomes a single point of failure for access enforcement. Most patterns recommend co-locating the broker with trusted workloads (e.g., inside the same namespace or cluster), and minimizing network exposure.

SPIFFE-based brokers can rely on attested identities and mutual TLS for agent-to-broker authentication, reducing the blast radius of compromised clients.

\subsection{Latency and Scale}

Adding a broker between the job and the resource introduces runtime latency. For most workloads, this overhead is negligible—but latency-sensitive tasks or high-frequency requests (e.g., serverless functions issuing credentials every second) may need caching, session delegation, or tiered trust models.

Designs should consider:
\begin{itemize}
  \item How long the credential is valid
  \item How often a job or workload requests access
  \item Whether credentials can be reused within a job lifecycle
\end{itemize}

\subsection{Auditability and Policy Lifecycle}

A core benefit of brokers is audit visibility: every issued credential has a policy trail and identity source. But this also means policy management becomes a critical surface. If policies are brittle, overly permissive, or hard to audit, the security posture can degrade.

Declarative policy engines like OPA or Cedar can help—but teams must still manage versioning, testing, and review processes for policy updates.

\subsection{Limitations and Future Directions}

Not all systems support SPIFFE or identity-aware credential issuance natively. Integrating brokers with legacy environments may require custom plugins or credential translation layers.

There are also open questions around revocation, chaining of justifications (e.g., who approved the access), and integrating human-in-the-loop workflows for sensitive actions.

These patterns lay the groundwork for richer policy enforcement, but maturity varies by ecosystem.

\section{Related Work}

The concept of separating identity from access is a foundational principle in modern zero trust architectures. Credential brokers extend this principle by enforcing runtime policy before issuing short-lived credentials. This work builds on emerging patterns in identity-aware infrastructure, federated identity, and policy-based access control.

SPIFFE and SPIRE define the workload identity primitives used throughout this paper \cite{spiffe-spec}. SPIRE enables runtime issuance of cryptographically verifiable identities (SVIDs) and supports attestation, federation, and trust bundle exchange across domains. While SPIRE is not a credential broker, it serves as the identity issuance layer upon which brokers can operate.

Credential brokering itself has been explored through commercial and open-source tools. Aembit provides one such broker platform that integrates with SPIFFE-issued identities, policy engines, and cloud APIs \cite{aembit}. Vault Agent templates and AWS STS federation mechanisms are also used as lightweight broker components in CI/CD systems \cite{aws-sts}. Snowflake’s adoption of workload IAM and brokered access demonstrates the operational benefits of credential brokers, including improved auditability and automation of access reviews~\cite{snowflake-aembit}.

Workload-to-workload identity and access management is gaining formal attention in IETF working groups like WIMSE and proposals for just-in-time authentication and tokenization mechanisms \cite{wimse}. These patterns aim to support least privilege, auditability, and secretless deployment—principles aligned with the Zero Standing Privilege (ZSP) model.

This paper contributes to the growing discussion around runtime policy enforcement for non-human actors. It builds upon ideas introduced in the companion work, \textit{Establishing Workload Identity for Zero Trust CI/CD: From Secrets to SPIFFE-Based Authentication} \cite{avirneni2025}, and focuses specifically on the architectural role of brokers in bridging identity issuance and access provisioning.

\section{Conclusion and Future Work}

Zero Trust principles—such as least privilege, continuous verification, and incident response—are essential for securing CI/CD pipelines in the face of modern threats~\cite{aptori-cicd}.

Credential brokers serve as a critical control point in CI/CD security—bridging the governance gap between identity issuance and access enforcement. By decoupling workload identity from access credentials, brokers enable just-in-time, policy-based access that aligns with Zero Trust principles.

In this paper, we examined the motivation behind credential brokers, explored core design patterns, and illustrated how SPIFFE-issued identities can integrate into broker workflows. This architecture supports dynamic credential issuance, policy-based authorization, and verifiable access audit across CI/CD pipeline stages.

While the patterns are promising, broker adoption requires careful integration with existing systems, thoughtful policy design, and operational maturity. Runtime brokers introduce tradeoffs in latency and complexity, but they offer scalable foundations for least privilege and intent-aware access control.

\subsection*{Next in the Series}

In the final paper of this series, we examine how runtime identity and policy signals can form a continuous control loop for compliance and security enforcement. We explore:

\begin{itemize}
  \item Credential lifecycle monitoring and revocation
  \item Real-time policy validation using justification tokens
  \item Feedback loops between brokers and authorization systems
\end{itemize}

This loop extends identity-aware access into a broader compliance and governance fabric, completing the Zero Trust CI/CD blueprint.

\appendix
\section*{Appendix}

\subsection*{A. Example Broker Policy (OPA)}
This policy follows ABAC principles and leverages OPA’s expressive policy language to enforce runtime access decisions~\cite{opa-abac}.
\begin{verbatim}
package authz

default allow = false

allow {
  input.spiffe_id == "spiffe://ci/org/deploy"
  input.resource == "s3://prod-release-artifacts"
  input.action == "write"
}
\end{verbatim}

This Rego policy governs access at the credential broker using SPIFFE-based identity.

\begin{itemize}
  \item \textbf{\texttt{package authz}}: Defines the policy namespace. All rules declared under this package will be evaluated under the \texttt{authz} context.
  \item \textbf{\texttt{default allow = false}}: Denies all requests unless explicitly permitted. This reflects a Zero Trust posture — no implicit access.
  \item \textbf{\texttt{input.spiffe\_id}}: Matches the SPIFFE ID of the requesting workload. This ensures only the intended CI job or workload receives access.
  \item \textbf{\texttt{input.resource}}: Specifies the resource being accessed (e.g., an S3 bucket). Resources can be matched by URI, label, or name.
  \item \textbf{\texttt{input.action}}: Defines the action being requested, such as "read", "write", or "deploy". This supports action-based authorization.
\end{itemize}

Together, these fields enable granular, identity-driven access control at runtime, enforced before credentials are issued. This policy could be extended to include time-based gating, environment metadata, or human approvals as additional conditions.

\subsection*{B. Sample Broker Flow Diagram}

    \begin{figure}[H]
      \centering
      \includegraphics[angle=270, width=0.5\textwidth]{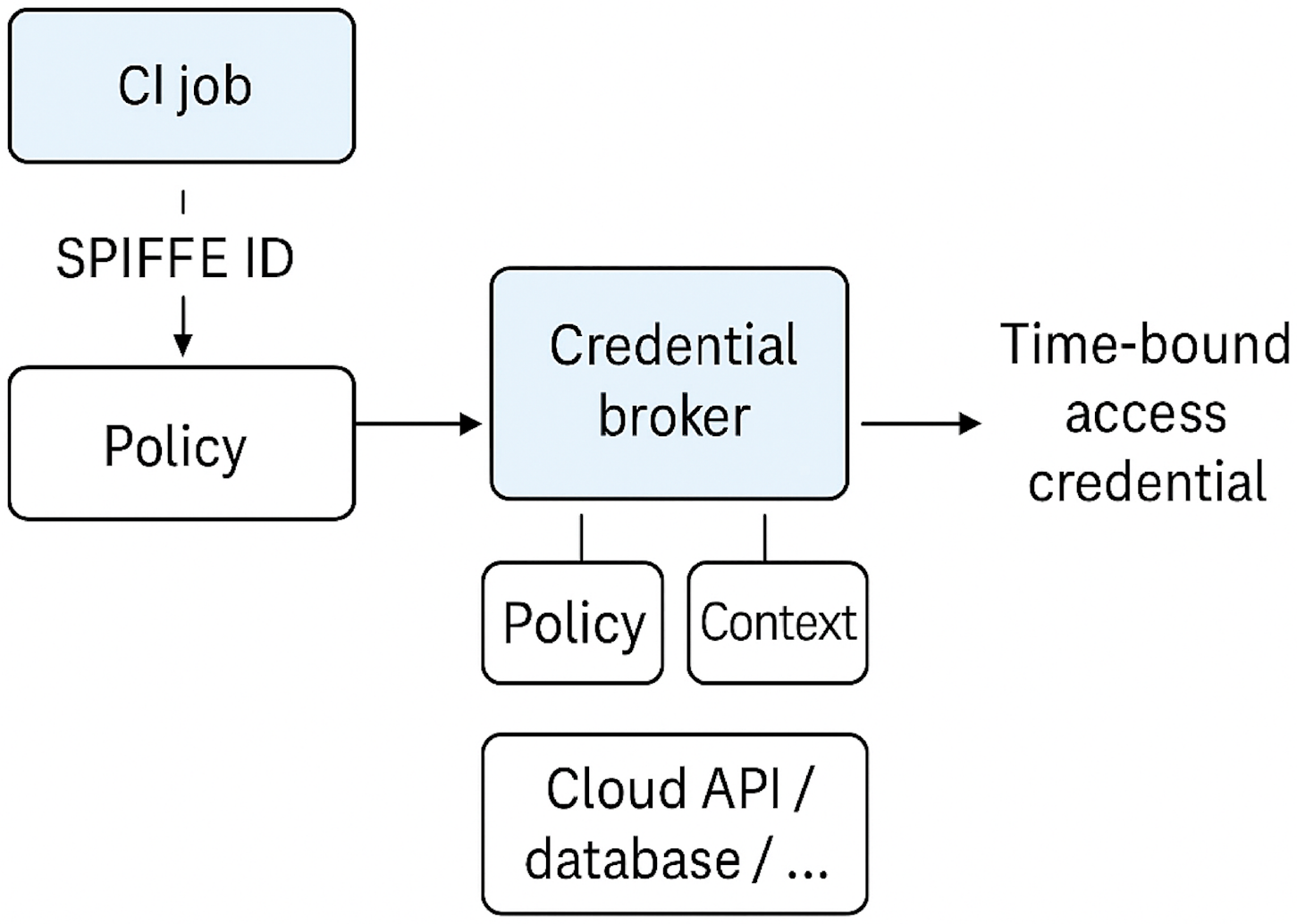}
      \caption{Reference architecture showing a SPIFFE-authenticated CI job interacting with a credential broker that evaluates policy and issues time-bound access credentials.}
      \label{fig:broker-architecture}
    \end{figure}

\end{document}